\documentclass[twocolumn,showpacs,amsmath,pra,aps,amssymb,superscriptaddress,nofootinbib]{revtex4-1} 
\usepackage[T1]{fontenc}
\usepackage[latin9]{inputenc}
\usepackage{times}
\usepackage{color} 
\usepackage{xspace}
\usepackage{amssymb,amsmath}
\usepackage{amsbsy}
\usepackage[pdftex]{graphicx}
\usepackage{bm}
\usepackage{float}
\usepackage{adjustbox}
\usepackage{epstopdf}

\usepackage[unicode,breaklinks]{hyperref}
\hypersetup{
    unicode=true,
    plainpages=false, 
    colorlinks=true,
    linkcolor=blue,
    citecolor=blue,
    filecolor=black,
    urlcolor=blue
}
\usepackage{url}
\usepackage{verbatim}

\begin{document}

\title{Coarsening and thermalisation properties of a quenched ferromagnetic spin-1 condensate}
\author{Lewis A. Williamson}  
\affiliation{Dodd-Walls Centre for Photonic and Quantum Technologies, Department of Physics, University of Otago, Dunedin 9016, New Zealand}
\author{P.~B.~Blakie}  
\affiliation{Dodd-Walls Centre for Photonic and Quantum Technologies, Department of Physics, University of Otago, Dunedin 9016, New Zealand}

\begin{abstract}
We examine the dynamics of a quasi-two-dimensional spin-1 condensate in which the quadratic Zeeman energy $q$ is suddenly quenched to a value where the system has a ferromagnetic  ground state. There are two distinct types of ferromagnetic phases, i.e.~a range of $q$ values where the magnetization prefers to be in the direction of the external field  (easy-axis), and a range of $q$ values where it prefers to be transverse to the field (easy-plane). We study the quench dynamics for a variety of $q$ values and show that there is a single dynamic critical exponent to characterize the scale invariant domain growth for each ferromagnetic phase.  For both quenches we give simple analytic models that capture the essential scale invariant dynamics, and correctly predict the exponents.  
Because the order parameter for each phase is different, the natures of the domains and the relevant topological defects in each type of coarsening  is also different. To explore these differences we characterize the fractal dimension of the domain walls, and the relationship of polar-core spin vortices to the domains in the easy-plane phase. Finally, we consider how the energy liberated from the quench thermalizes in the easy-axis quench.  We show that local equilibrium is established in the spin waves on moderate time scales, but continues to evolve as the domains anneal.\end{abstract}

\maketitle

\section{Introduction}

After a rapid quench through a symmetry-breaking phase transition a many-body system will form causally disconnected spatial domains, each making an independent choice for the symmetry broken order parameter. The coarsening dynamics of how such a system subsequently evolves towards equilibrium is an area of broad interest \cite{Bray1994}. At long times after the quench a universal scaling regime can develop: correlation functions of the order parameter collapse to a universal scaling function (independent of time $t$) when space is scaled by a characteristic length $L(t)$. The growth law for this characteristic length $L(t)\sim t^{1/z}$ yields the dynamic critical exponent $z$.

While most classical theories for coarsening dynamics have been developed for dissipative models related to temperature quenches, recently there has been growing interest in the dynamics of systems under conservative Hamiltonian evolution, particularly due to developments with ultra-cold atomic gases \cite{Damle1996a,zheng2000,koo2006,asad2007,Takeuchi2012a,Hofmann2014}. 
Here we will focus on the coarsening dynamics of a ferromagnetic spin-1 condensate.  Such spinor condensates are unique in that they exhibit both superfluid and magnetic order \cite{Kawaguchi2012R,StamperKurn2013a}, and have a rich set of zero temperature phases which can be conveniently explored in experiment. A motivating experiment was performed by the Berkeley group with a $^{87}$Rb condensate that was quenched from a non-magnetized (polar) phase into a ferromagnetic phase by a sudden change in the quadratic Zeeman energy ($q$) of the atoms \cite{Sadler2006a} (also see \cite{Higbie2005a,Leslie2009a,Bookjans2011b,Guzman2011a,De2014a}). More generally, depending on the value of $q$ quenched to, the magnetization that develops can have easy-axis or easy-plane symmetry. In previous work \cite{Williamson2016a} we numerically demonstrated that the late time coarsening in these two cases are described by the hydrodynamic binary fluid and model E dynamic universality classes, respectively. 

In this paper we expand upon our earlier work, and provide a fuller description of the coarsening dynamics of the ferromagnetic spin-1 condensate. We simulate quenches for a wide range of $q$ values in the easy-axis and easy-plane regimes, and demonstrate that the exponents obtained are universal. Supporting these results we develop models of the key processes governing coarsening and apply scaling arguments to obtain the same exponents.
We consider the mean value and fluctuations of the magnetization in the post-quench system to reveal the thermalization of the energy liberated by the quench. For the easy-axis case we develop a scheme for thermometry using spin-waves and demonstrate that these modes thermalize on a much faster time scale than the order parameter evolution governed by the coarsening dynamics. We also consider the domain wall structure by evaluating the order parameter structure factor, where the domain wall properties are revealed by a Porod tail feature. Interestingly the analysis of the Porod tail for the easy-plane case  suggests that the domain walls have a fractal structure. We verify this by directly applying a box-counting algorithm to the spatial domains of the coarsened system. The easy-plane phase has an order parameter that supports vortices (polar-core vortices) as topological defects. We evaluate the number of topological defects during the coarsening evolution and show that this is directly related to the coarsening length scale of domains.  

This work we report here provides a thorough analysis of the coarsening dynamics for the ferromagnetic spin-1 condensate, and establishes a firm basis and set of tools for future work on other quenches (e.g.~temperature quenches) and for work on the anti-ferromagnetic and higher spin cases.

The outline of this paper is as follows. In Sec. II we introduce the Gross-Pitaevskii formalism, and the relevant order parameters and their symmetries for the two phase transitions we explore. We also outline the numerical methods we use to simulate the phase transition dynamics. The main results are presented in Sec. III. We begin by examining the growth of local order following a quench in the quadratic Zeeman energy. We then introduce the order parameter correlation functions and examine the nature of dynamic scaling in the post-quench coarsening dynamics. We examine the role of vortices in the easy-plane quench and also examine the fractal dimension of domain boundaries in quenches via the order parameter structure factor and by applying a box-counting algorithm directly to the domains. For both phase transitions we develop analytic models for the relevant degrees of freedom during the coarsening regime. Dimensional analysis of these models yields the dynamic critical exponents found in the simulations. Finally, we examine the thermalisation that occurs in the easy-axis quench. We find that spin-waves in the easy-axis system thermalise on a rapid timescale compared to the order parameter coarsening dynamics. Finally, in Sec. IV we conclude and discuss the outlook for future work in this area.

\section{Formalism}
\subsection{The Spin-1 Gross-Pitaevskii equations} 
The system we consider is a homogeneous quasi-two-dimensional (quasi-2D) spin-1 condensate  
described by the Hamiltonian~\cite{Ho1998a,Ohmi1998a}
\begin{align}\label{spinH}
H\!=\!\int\!d^2\bm{x}\left[\bm{\psi}^\dagger\!\left(\!-\frac{\hbar^2\nabla^2}{2M}-pf_z+qf_z^2\right)\!\bm{\psi}+\frac{g_n}{2}n^2+\frac{g_s}{2}\left|\bm{F}\right|^2\right]\!.
\end{align}
Here $\bm{\psi}\equiv (\psi_{1},\psi_0,\psi_{-1})^T$ is a three component spinor describing the condensates in the three spin levels and $p$ and $q$ are, respectively, the linear and quadratic Zeeman shifts arising from the presence of an external field along $z$.  The term $g_nn^2$ describes the density interaction with coupling constant $g_n$, where  $n\equiv\bm{\psi}^\dagger\bm{\psi}$ is the areal density. 
The term $g_s|\bm{F}|^2$ describes the spin interaction with coupling constant $g_s$, where  $\bm{F}\equiv \bm{\psi}^\dagger\bm{f}\bm{\psi}$  is the areal spin density for the spin-1 matrices $(f_x,f_y,f_z)\equiv\bm{f}$. 
For the system to be mechanically stable we require $g_n>0$. The coupling $g_s$ can be positive or negative resulting in anti-ferromagnetic or ferromagnetic interactions, respectively.  Here we consider the case of ferromagnetic interactions, i.e.~ $g_s<0$, as realized in $^{87}$Rb  condensates \cite{Chang2004a}. In spinor condensate experiments the quasi-2D regime has been realized by using a trapping potential with tight confinement in one direction (e.g.~see \cite{Sadler2006a}). Our interest is in homogeneous systems where the phase transition dynamics are simpler, noting that recent experiments have realized flat-bottomed optical traps for this purpose \cite{Navon2015a,Chomaz2015a} (also see~\cite{damle1996b}).

The dynamics of the system can be described by the three coupled Gross-Pitaevskii equations (GPEs),
\begin{align}\label{spinGPEs}
i\hbar\frac{\partial\bm{\psi}}{\partial t}=\left(-\frac{\hbar^2\nabla^2}{2M}-pf_z+qf_z^2+g_nn+g_s\bm{F}\cdot\bm{f}\right)\bm{\psi}.
\end{align}
The linear Zeeman term can be removed by moving to a rotating frame, $\bm{\psi}\rightarrow e^{ipf_zt/\hbar}\bm{\psi}$, so that from hereon we set $p=0$.

\subsection{Phase diagram, symmetries and conservation laws}\label{SecPDSymmetries}

\begin{figure}
\centering
\includegraphics[width=0.5\textwidth]{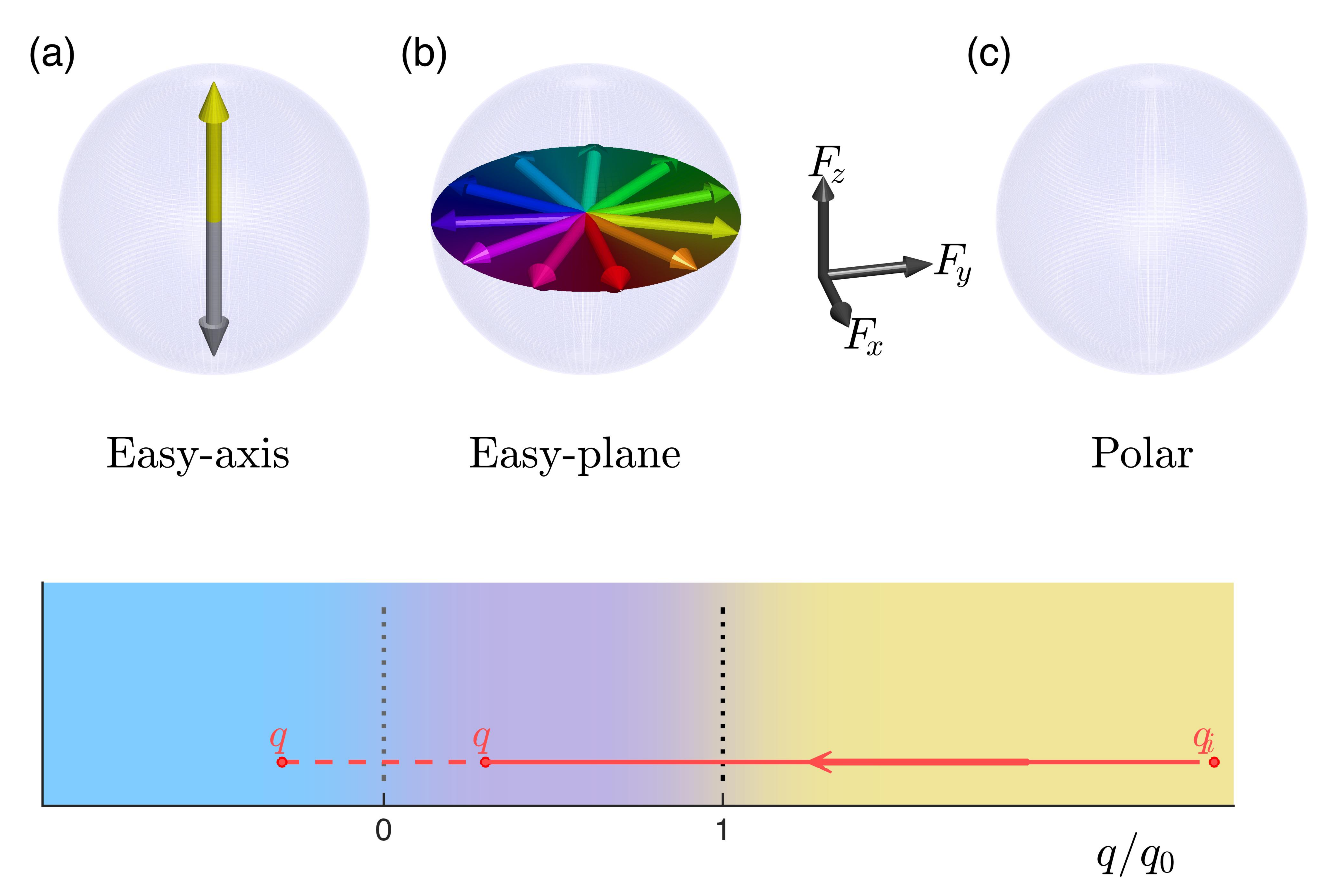}
\caption{\label{phaseDiag}Phase diagram and ground states of a spin-1 condensate for $p=0$. The spheres show the direction of magnetization in the three states. (a) For $q<0$ the magnetization lies along the $F_z$ axis and the state is termed \emph{easy-axis}. (b) For $0<q<q_0$ the magnetization lies in the transverse ($F_x$-$F_y$) plane and the state is termed \emph{easy-plane}. (c) For $q>q_0$ the $m=\pm 1$ levels are unoccupied and the system is unmagnetized. This state is termed \emph{polar}.}
\end{figure}

The Hamiltonian~\eqref{spinH} gives rise to different magnetic ground states depending on the value of $q$~\cite{Kawaguchi2012R}. It is convenient to express these ground states in the form  $\bm{\psi}=e^{i\theta}\sqrt{n_0}\bm{\xi}$, where  $n_0$ is the (uniform) condensate density, $\theta$ is a global phase, and $\bm{\xi}$ is a normalized spinor (i.e.~$\bm{\xi}^\dagger\bm{\xi}=1$). Here our primary interest is in the spin order that develops in the system, and hence in the properties and symmetries of $\bm{\xi}$. A schematic phase diagram and representation of the ground states important to this paper are shown in Fig.~\ref{phaseDiag}. There are two ferromagnetic phases (a) and (b), which differ in their symmetries, and a non-magnetized polar phase (c).

\textit{Easy-axis} ferromagnetic ground state (a) has the normalized spinor
\begin{align}
\bm{\xi}_{\mathrm{EA}}=\left(\begin{array}{c}1\\0\\0\end{array}\right)\quad\mbox{or}\quad\left(\begin{array}{c}0\\0\\1\end{array}\right),
\end{align}
with a magnetization of +1 or -1 along $z$, respectively. This state is degenerate under reflections in the $x$-$y$ plane, giving rise to a $\mathbb{Z}_2$ manifold of ground states. This state is the ground state of the system for $q<0$.

\textit{Easy-plane} ferromagnetic ground state (b) has 
\begin{align}
\bm{\xi}_{\mathrm{EP}}=\frac{1}{2}\left(\begin{array}{c} e^{-i\varphi}\sqrt{1-q/q_0}\\ \\ \sqrt{2(1+q/q_0)}\\ \\ e^{i\varphi}\sqrt{1-q/q_0} \end{array}\right),\quad 0\le\varphi<2\pi,
\end{align}
where we have defined $q_0=2|g_s|n_0$. This state has a magnetization of length $F=n_0\sqrt{1-(q/q_0)^2}$ lying in the $x$-$y$ plane, at an angle $\varphi$ to the $x$-axis. It is degenerate under rotations about the $F_z$ axis, giving rise to a $\text{SO}(2)$ manifold of ground states. The easy-plane phase is the ground state for $0<q<q_0$, with a critical point at $q=q_0$ separating it from the polar phase.

\textit{Polar}  ground state (c) has 
\begin{align}
\bm{\xi}_{\mathrm{P}}=\left(\begin{array}{c}0\\1\\0\end{array}\right),
 \end{align}
with all components of the spin density being zero.

Evolution under Eq.~\eqref{spinGPEs} preserves $\mathbb{Z}_2$ symmetry corresponding to reflections in the transverse plane and $\text{SO}(2)$ symmetry corresponding to rotations about the $z$ axis. Therefore an initial state adhering to these symmetries will maintain these symmetries. For the ferromagnetic states (a) and (b), a symmetry in the system is broken when the system chooses a ground state within the ground state manifold. The nature of the ground state manifold determines this symmetry. In the easy-axis phase the order parameter is $\propto F_z$ and a choice of ground state breaks the $\mathbb{Z}_2$ symmetry. In the easy-plane phase the order parameter is $\propto (F_x,F_y)$ and a choice of ground state breaks the $\text{SO}(2)$ symmetry. The order parameter in the easy-axis phase is conserved under the evolution of~\eqref{spinGPEs}, whereas the order parameter in the easy-plane phase is not. The symmetry and conservation properties of the order parameter determine the critical behaviour of the transition to each phase.

\subsection{Details of simulation method}\label{SecSimDetails}
In this work we consider the phase ordering dynamics after a quench of the quadratic Zeeman energy from an initial value $q_i>q_0$ where the ground state is polar  to a final value $q<q_0$, as shown schematically in Fig.~\ref{phaseDiag}. The system will then order into either the easy-axis phase (for $q<0$) or the easy-plane phase (for $0<q<q_0$). We simulate the dynamics using the GPEs (\ref{spinGPEs}) with noise added to the initial state to seed the growth of symmetry breaking domains. 

For the initial state we take
\begin{align}
\bm{\psi}(\bm{x})=\sqrt{n_0}\left(\begin{array}{c}0\\1\\0\end{array}\right)+\bm{\delta}(\bm{x})
\end{align}
where $\sqrt{n_0}\left(0,1,0\right)^T$ is the polar condensate wavefunction and $\bm{\delta}$ is a small noise field given by
\begin{align}
\bm{\delta}(\bm{x})=\sum_{\bm{k}}\left(\begin{array}{c}\alpha^+_{\bm{k}}e^{i\bm{k}\cdot\bm{x}}\\ \\  \alpha^0_{\bm{k}}u_ke^{i\bm{k}\cdot\bm{x}}-{\alpha^{0*}_{\bm{k}}}v_ke^{-i\bm{k}\cdot\bm{x}}\\ \\
\alpha^-_{\bm{k}}e^{i\bm{k}\cdot\bm{x}}\end{array}\right).
\end{align} 
Here the $\{\alpha^m_{\bm{k}}\}$ are independent complex Gaussian random variables with 
\begin{align}
\langle \alpha^{m'*}_{\bm{k}'}\alpha^{m}_{\bm{k}}\rangle=\frac{1}{2}\delta_{mm'}\delta_{\bm{k}'\bm{k}},
\end{align}
and we take $\alpha^0_\mathbf{0}=0$ to omit adding noise to the condensate mode. The amplitudes $\{u_k,v_k\}$ are given by
\begin{align}
u_k=\sqrt{ \frac{\epsilon_k+g_nn_0}{2\sqrt{\epsilon_k\left(\epsilon_k+2g_nn_0\right)}}-\frac{1}{2}},\quad v_k=\sqrt{u_k^2-1},
\end{align}
with $\epsilon_k=\hbar^2k^2/2M$. The noise added this way corresponds to adding a half-quantum of occupation to the Bogoliubov modes for the polar phase at large $q_i$~\cite{Lamacraft2007a} as per the truncated Wigner presciption \cite{cfieldRev2008}.

In  experiments with $^{87}$Rb the spin interaction is much weaker than the density interaction with $g_n/|g_s|\sim100$~\cite{Kawaguchi2012R}. To observe universal dynamics we must simulate our system over many spin times $t_s\equiv \hbar/2|g_s|n_0$. With large density interaction the system has to resolve fast but largely unimportant density fluctuations. This slows down the numerics substantially. To allow faster simulations, we use the more moderate ratio of interaction parameters $g_n/|g_s|=10$. We have also run simulations with $g_n/|g_s|=3$ and obtained consistent results \cite{Williamson2016a}. Density fluctuations may add noise to order parameter correlations for a single simulation, but results obtained by averaging over simulations should remove this. We expect little change in the phase ordering dynamics for higher interaction parameter ratios, which would reduce the density fluctuations leading to less noise in single simulations.

We use a  condensate density of  $n_0=10^4/\xi_s^2$, where $\xi_s\equiv\hbar/\sqrt{2|g_s|n_0 M}$ is the spin healing length. To numerically evolve the GPEs we represent each component of the spinor field $\bm{\psi}$ on a 2D square grid with dimensions $l\times l$ covered by an $N\times N$ grid of equally spaced points. For simulations of quenches to the easy-axis phase we use grids of size $l=800\xi_s$ with $N=1024$ points. For the easy-plane cases we use $l=1600\xi_s$ with $N=2048$ points.  We evolve the spin-1 GPEs~\eqref{spinGPEs} using an adaptive step Runge-Kutta method that uses Fast Fourier transforms to evaluate the kinetic energy operators with spectral accuracy.  The quadratic Zeeman energy is set at the final quench value $q<q_0$ for the duration of the simulation dynamics, so that the quench is effectively instantaneous at $t=0$.

\section{Results and Analysis}
\subsection{Post-quench growth of magnetization}
\begin{figure*}
\includegraphics[width=0.8\textwidth]{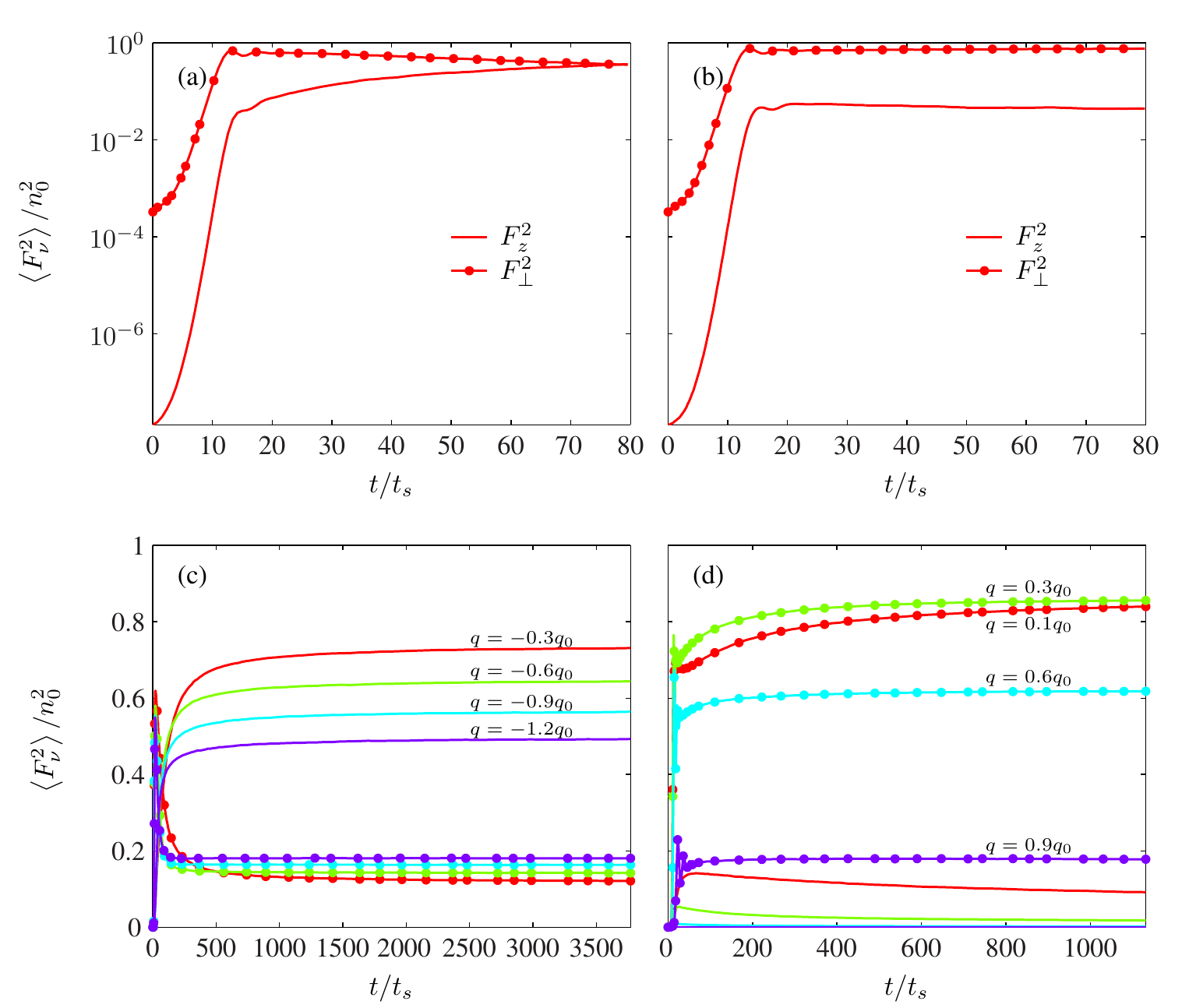}
\caption{\label{Allmag} Growth of magnetization following a quench of a spin-1 condensate. Plain lines correspond to $F_z$ magnetization; lines with dots correspond to $F_\perp$ magnetization. Short-time dynamics for a quench to the (a) easy-axis phase ($q=0.3 q_0$)  and (b) easy-plane phase ($q=-0.3q_0$). Long-time dynamics for the (c) easy-axis phase and the (d) easy-plane phase. In all cases the (local) magnetization is calculated as a spatial average over the system at each time, i.e.~$\langle F_\nu^2\rangle= l^{-2}\int d^2\mathbf{r}\,F^2_\nu(\mathbf{r}) $ for $\nu=z,\perp$. In (c), red data is for $q=-0.3q_0$, green data is for $q=-0.6q_0$, light blue data is for $q=-1.2q_0$ and purple data is for $q=-1.2q_0$. In (d), red data is for $q=0.1q_0$, green data is for $q=0.3q_0$, light blue data is for $q=0.6q_0$ and purple data is for $q=0.9q_0$.}
\end{figure*}

Following the quench to either the easy-axis or easy-plane phase, the system develops local magnetization (i.e.~the spin density becomes non-zero). The development of the magnitude of the transverse $\mathbf{F}_\perp=(F_x,F_y)$ and longitudinal $F_z$ magnetization is shown in  Fig.~\ref{Allmag}. 
 The initial growth is exponential \cite{Saito2007a,Barnett2011}  (also see \cite{Lamacraft2007a,Saito2007b,Uhlmann2007a,Damski2007a}) and is similar for quenches to values of $q$ in the easy-axis and easy-plane regimes [see Figs.~\ref{Allmag}(a),(b)]. The exponential growth ceases after a time $t\sim 10t_s$. On longer time scales the magnetization develops an easy-axis  [i.e.~$F_z$ dominates,   Fig.~\ref{Allmag}(c)] or easy-plane [i.e.~$F_{\perp}$ dominates,  Fig.~\ref{Allmag}(d)] character, revealing the preferred order of the phase the system has been quenched into.
The magnetization reaches a steady magnitude after a time $t\sim 200 t_s$.

\begin{figure*}
\includegraphics[width=0.8\textwidth]{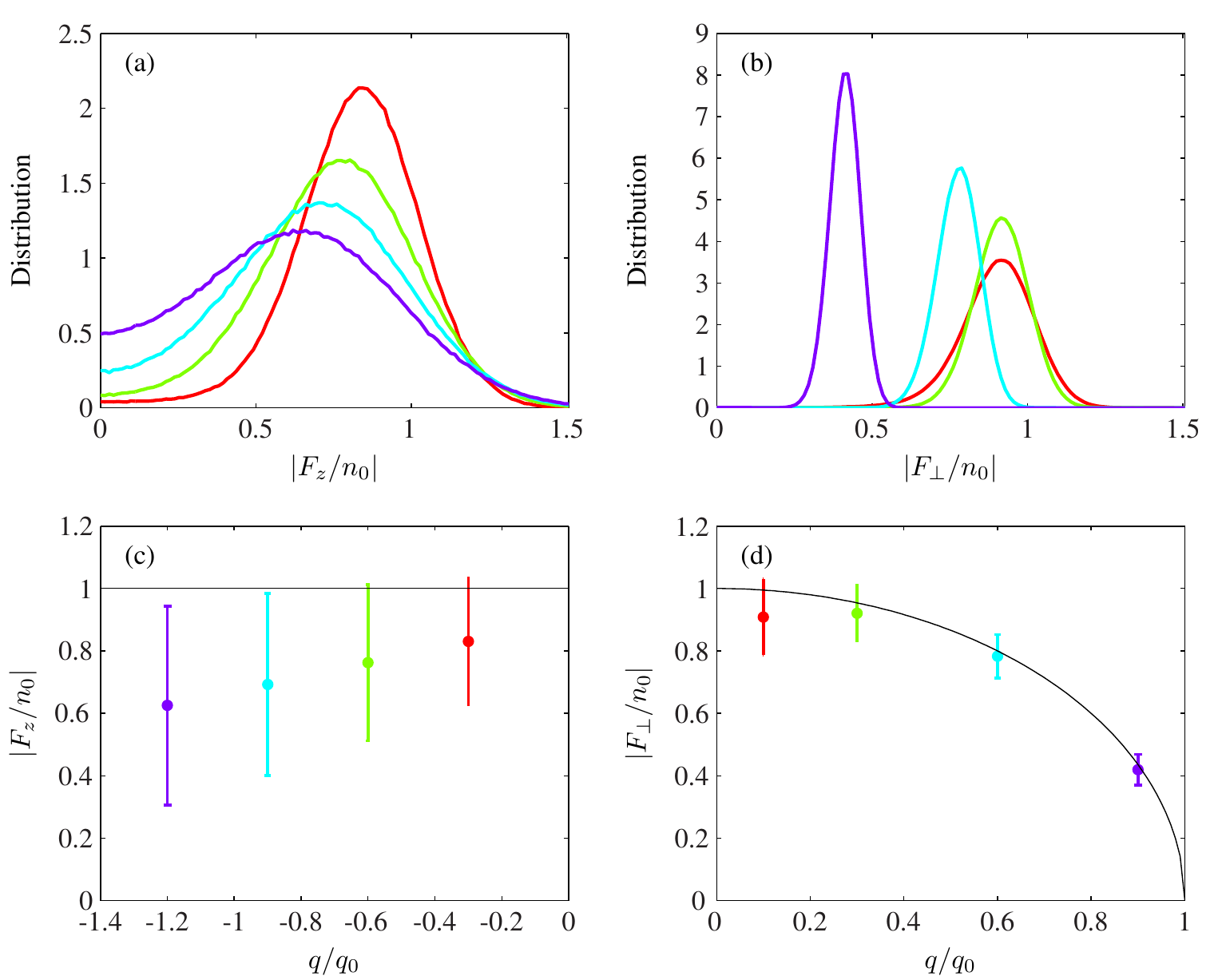}
\caption{\label{Allfluc} Normalized distribution of magnetization long after the quench. The distribution of the (a) longitudinal magnetization for easy-axis quenches and (b) transverse magnetization for easy-plane quenches. These results are evaluated by making a histogram of the magnetization magnitudes sampled over space. In (a), red (dot-dashed) data is for $q=-0.3q_0$, green (dotted) data is for $q=-0.6q_0$, light blue (dashed) data is for $q=-1.2q_0$ and purple (solid) data is for $q=-1.2q_0$. In (b), red (dot-dashed) data is for $q=0.1q_0$, green (dotted) data is for $q=0.3q_0$, light blue (dashed) data is for $q=0.6q_0$ and purple (solid) data is for $q=0.9q_0$.
In (c) and (d) the mean  value and spread (standard deviation) obtained from the results in (a) and (b) are compared to the ground state magnetization (solid black lines). We note that for the easy-plane phase the ground state transverse magnetization depends on $q$ as $|F_\perp|=n_0\sqrt{1-(q/q_0)^2}$. Results (a), (c) are calculated at $t=3770t_s$, while (b), (d) are calculated at $t=1131t_s$ .}
\end{figure*}

Immediately after the quench, the system is still in the polar phase and so is out of equilibrium. This gives the system an energy in excess of that of the ground state. The excess energy can be calculated from \eqref{spinH} yielding
\begin{align}\label{Eex}
\Delta E = \left\{\begin{array}{lll}\left(\frac{1}{4}q_0-q\right)n_0l^2, & \quad & q<0, \\ \\ \frac{1}{4}q_0\left(1-{q}/{q_0}\right)^2n_0l^2, & & 0<q<q_0.\end{array}\right. 
\end{align} 
This excess energy is available for thermalization and heats the system. This results in fluctuations in the magnetization [Fig.~\ref{Allfluc}(a),(b)] and a reduction in the magnitude of the magnetization from the ground state value [Fig.~\ref{Allfluc}(c),(d)]. Both effects are more pronounced for deeper quenches (i.e. to lower $q$ values) which have higher values of $\Delta E$. 
%

\subsection{Universal coarsening dynamics}
\begin{figure*}[t]
\centering
\includegraphics[width=\textwidth]{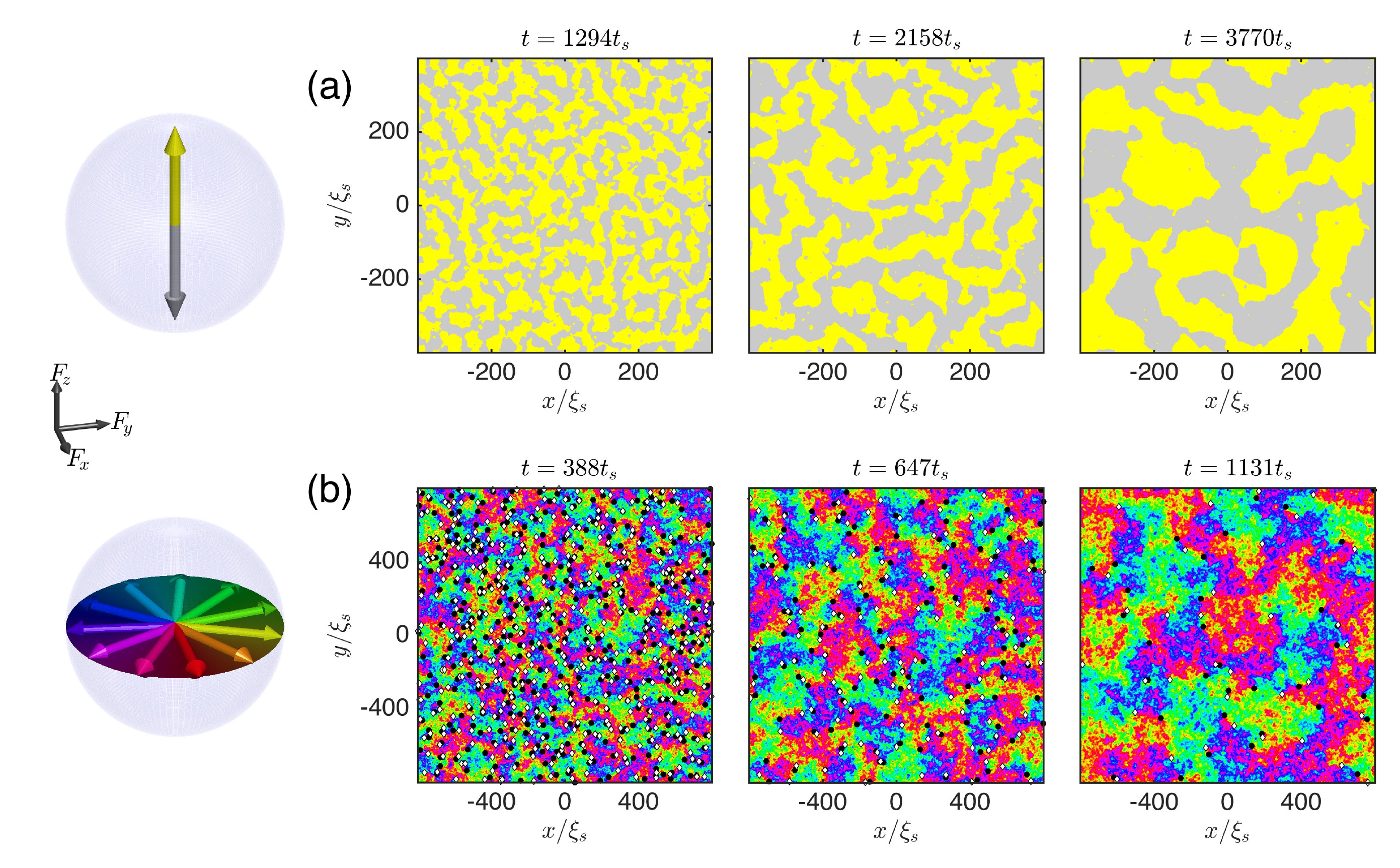}
\caption{\label{domain}Growth of domains for (a) the easy-axis phase and (b) the easy-plane phase. Colour scales are indicated by the respective spin spheres. Positive (black circles) and negative (white diamonds) polar-core spin vortices are present in the easy-plane system. Vortex-antivortex annihilation accompanies the growth of domains in this phase.}
\end{figure*}

The magnetization dynamics clearly show that the longitudinal (transverse) magnetization dominates at times sufficiently long after the quench for the easy-axis (easy-plane) quenches. As discussed in Sec.~\ref{SecPDSymmetries} this motivates us to define the order parameters
\begin{align}
\phi(\mathbf{r})&=\frac{1}{n_0}F_z(\mathbf{r}),\hspace{0.12cm}&\mbox{Easy-axis},\hspace{0.22cm}\\
\phi(\mathbf{r})&=\frac{1}{n_0}\mathbf{F}_\perp(\mathbf{r}),\hspace{0.12cm}&\mbox{Easy-plane},
\end{align}
to characterize the development of order. Crucially, our interest lies not in the emergence of local order (as characterized by the local magnetization), but in the evolution of the ordered domains on long time scales. Examples of these domains and their evolution are shown in Fig.~\ref{domain} revealing the tendency of the domains to grow with time. These domains are described by the order parameter correlation function
\begin{align}\label{Gdef}
G(r,t)=\langle \phi(\mathbf{0})\cdot \phi(\mathbf{r})\rangle_t,
\end{align}
where the average is taken at a time $t$ after the quench. In practice we can calculate this correlation function utilizing the translational invariance of our simulations to spatially average  i.e.~calculate
\begin{align}
G(\mathbf{r},t)=\frac{1}{l^2} \int d^2\mathbf{r}' \phi(\mathbf{r}')\cdot \phi(\mathbf{r}'\!+\!\mathbf{r}),
\end{align} 
and also use isotropy to perform an angular average over all points at a distance $r$. To further improve statistical sampling we also average over 8 simulation trajectories conducted with different initial noise.

\begin{figure*}
\includegraphics[width=0.8\textwidth]{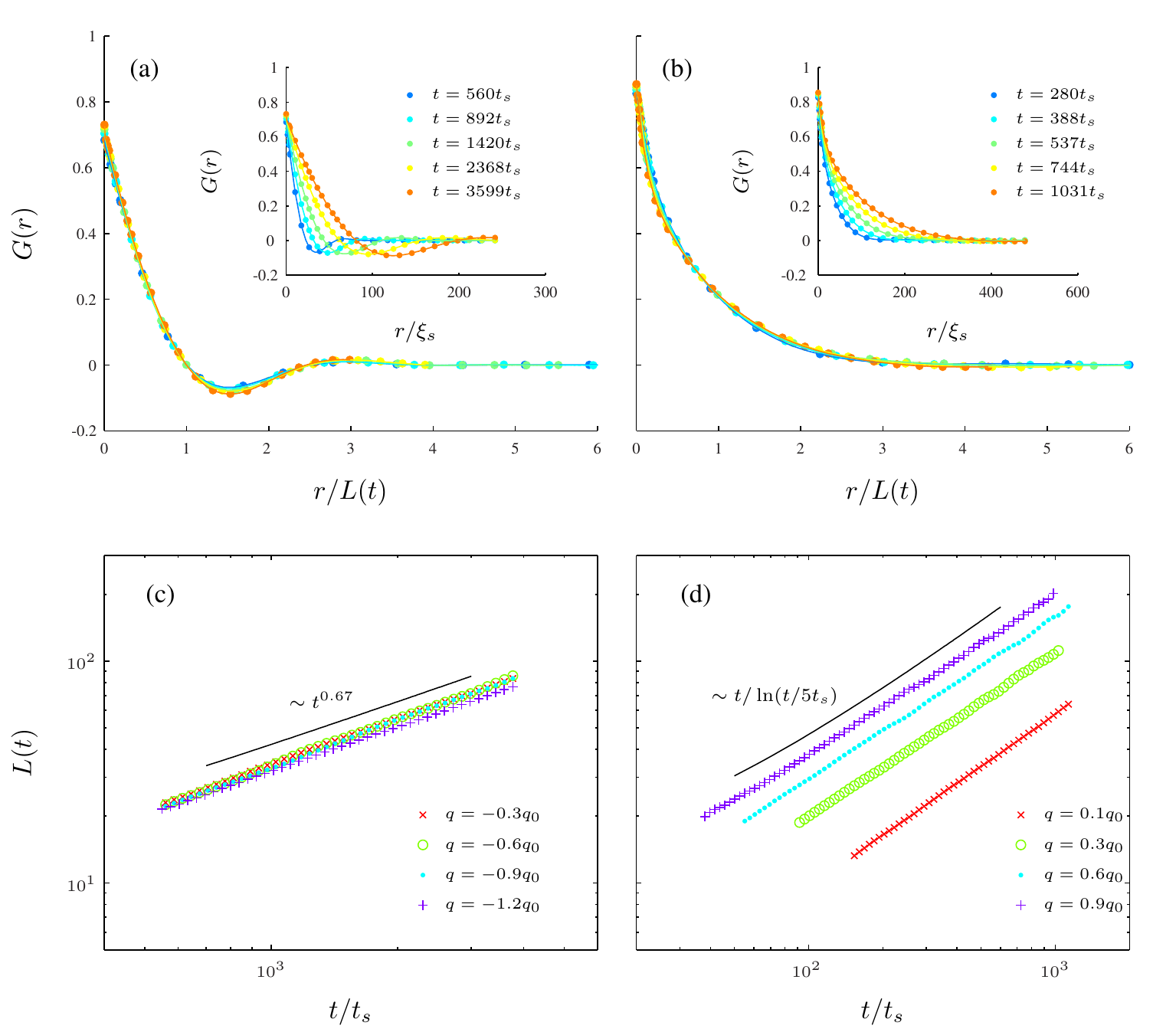}
\caption{\label{GLfig} Collapse of the order parameter correlation function onto a single curve when lengths are scaled by the growing length scale $L(t)$ for (a) the easy-axis phase and (b) the easy-plane phase. The insets show the correlation functions before rescaling. (c) and (d) show the length scales $L(t)$ used for the correlation function collapse in (a) and (b), respectively. Solid lines indicate power law growths with $1/z=0.67$ and $1/z=1$, respectively. Data in (a) is for $q=-0.3$ and data in (b) is for $q=-0.3$.  
}
\end{figure*}

The temporal evolution of the correlation function is shown in the insets to Figs.~\ref{GLfig}(a) and~(b). The length scale over which the correlation function decays can be taken to define a characteristic domain size. As time progresses this length scale is seen to grow as order extends over large regions.  As anticipated by the theory of phase ordering kinetics, we find that this growth exhibits dynamic scale invariance: Correlations of the order parameter at late times collapse onto a single universal curve $f(r)$ when lengths are scaled by a large characteristic length scale $L(t)$, i.e.~
\begin{align}
f(r)=G(r/L(t),t).
\end{align}
For the easy-axis phase, we take $L(t)$ to be the first zero crossing of $G(r,t)$. For the easy-plane phase, we take $L(t)$ to be the point where $G(r,t)=0.25G(0,t)$.
Using these length scales we demonstrate the correlation function collapse in Figs.~\ref{GLfig}(a) and~(b).  

The growths of $L(t)$ for the easy-axis quenches are shown in  Fig.~\ref{GLfig}(c). Here we find that for a range of $q$ values $L(t)\sim t^{1/z}$ with $z=3/2$. This is consistent with the dynamic critical exponent of a binary fluid in the inertial hydrodynamic regime~\cite{Furukawa1985}. 
The scale invariant dynamics can be ascribed to a process of hydrodynamic flow of the $F_z$ superfluid velocity, see Sec.~\ref{SecEAmodel}. 
 
\begin{figure}
\includegraphics[width=0.5\textwidth]{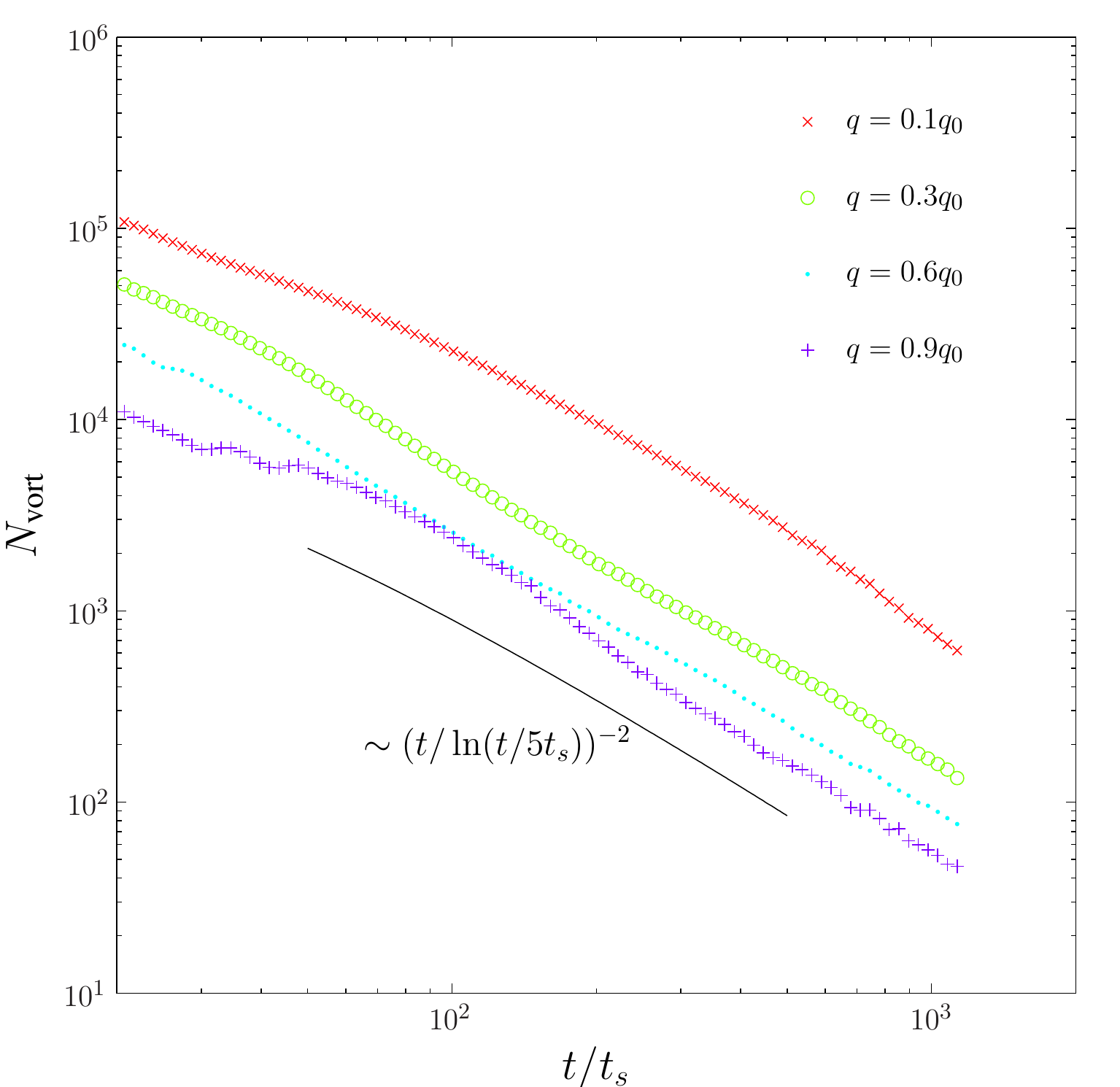}
\caption{\label{Lvfig} Number of polar-core vortices $N_\text{vort}$ (irrespective of sign) versus time in easy-plane quenches. As the domains grow, the number of vortices decays as $L(t)^{-2}$ [indicated as black line using fit for $L(t)$ from Fig.~\ref{GLfig}(d)] }
\end{figure}

In the easy-plane phase, we find that $L(t)\sim t/\ln(t/5t_s)$ for a range of $q$ values, giving a dynamic critical exponent of $z=1$ with a logarithmic correction, see Fig.~\ref{GLfig}(d). A dynamic critical exponent of $z=1$ is consistent with the Model E universality class, which describes a 2D non-conserved order parameter coupled to a second conserved field. This fits our system well, where the second conserved field is $F_z$, see Sec.~\ref{SecEPmodel} and~\cite{Lamacraft2007a}. The logarithmic correction is included to account for the presence of vortices, in analogy with the $XY$-model. Much work has shown that vortices in the $XY$-model slow the rate of coarsening and give rise to a logarithmic correction to scaling~\cite{Yurke1993a,bray1994b,rutenberg1995b,lee1995,Rojas1999a,Bray2000a,berthier2001,jelic2011} (see also~\cite{ching2001,Forrester2013a}) so that true dynamic scale invariance $L(t)\sim t^{1/z}$ is only obtained after a very long time. In the easy-plane phase the order parameter supports polar-core vortices. These vortices consist of a phase winding in the (in-plane) magnetization around an unmagnetized core. The state of a polar-core vortex is 
\begin{align}\label{pcvstate}
\bm{\psi}_{\mathrm{vort}}=\sqrt{\frac{n}{2}}\left(\begin{array}{c}\sin\beta e^{-i\theta}  \\ \sqrt{2}\cos\beta  \\ \sin\beta e^{i\theta}\end{array}\right),
\end{align}
 where far from the vortex core $\cos\beta=\sqrt{(1+q/2|g_s|n)/2}$ \cite{Kawaguchi2012R}. The in-plane magnetisation angle $\theta$ rotates by $2\pi\kappa$ ($\kappa\in\mathbb{Z}$) around the vortex centre. These vortices are known as polar-core vortices because the particle density is in the $\psi_0$ component at the centre of the vortex, to avoid the phase singularity in the $\psi_{\pm 1}$ components. We only observe polar-core vortices of charge $\kappa=\pm 1$ during the coarsening regime, noting that higher values of $|\kappa|$ are unstable. We find that the domain growth is associated with the annihilation of these vortices [see Fig.~\ref{domain}(b)], and that the number of vortices is correlated with $L(t)$. As $L(t)$ grows, the density of vortices and therefore the total vortex number decay as $L(t)^{-2}$, see Fig.~\ref{Lvfig}.

A phase winding of the spin angle $\theta$ corresponds to a circulation of the $F_z$ superfluid current, since $F_z \propto \nabla\theta$~\cite{Yukawa2012}. It is feasible to also have circulation in mass and other spin currents, which would give rise to other types of vortices. We only observe polar-core spin vortices during the coarsening regime (see also~\cite{Kudo2015a}).

We note that the exponents obtained in Fig.~\ref{GLfig}(a) and (d) vary slightly with $q$. This is consistent with finite-size effects~\cite{huse1986,Hofmann2014} and statistical sampling over the time range we can simulate. For the easy-axis quench we obtain a range of $1/z=0.66-0.70$. For the easy-plane quench we obtain a range of $1/z=0.98-1.01$. A fit of the form $L(t)~t^{1/z}$ also fits the numerical data in Fig.~\ref{GLfig}(d) well, and yields $1/z=0.71-0.78$, depending on $q$. To differentiate this fit from one with $z=1$ and a logarithmic correction would require simulation times $t>10^4t_s$, which is well beyond the range of our simulations. Furthermore, including a logarithmic correction in the easy-plane quench shrinks the range of exponents obtained to $1/z=0.98-1.01$, consistent with universal scaling.

\subsection{Fractal dimension of domains}
The order parameter domains in Fig.~\ref{domain} are separated by domain boundaries. By examining the correlation function~\eqref{Gdef} at length scales $r<L(t)$, we are able to extract information about these boundaries. In particular, we can determine the dimension of the domain boundaries, which can be non-integer if the boundary has a fractal structure. It can be shown that the small $r/L$ behaviour of the correlation function behaves as~\cite{bale1984}
\begin{align}\label{bccorr}
1-G(r,t)\sim \left(\frac{r}{L(t)}\right)^{D-D_b}
\end{align}
where $D_b$ is the fractal dimension of the domain boundary and $D$ is the system dimension. For the smooth case in two dimensions, i.e. $D=2$, $D_b=1$, Eq.~\eqref{bccorr} reflects the intuitive result that the probability that two points a distance $r$ apart will lie in opposite domains is $\sim r/L$ for $r\ll L$. In general, the probability that two points a distance $r$ apart will lie in opposite domains is $\sim (r/L)^{D-D_b}$ for $r\ll L$ from which one can derive Eq.~\eqref{bccorr}.

\begin{figure}
\includegraphics[width=0.5\textwidth]{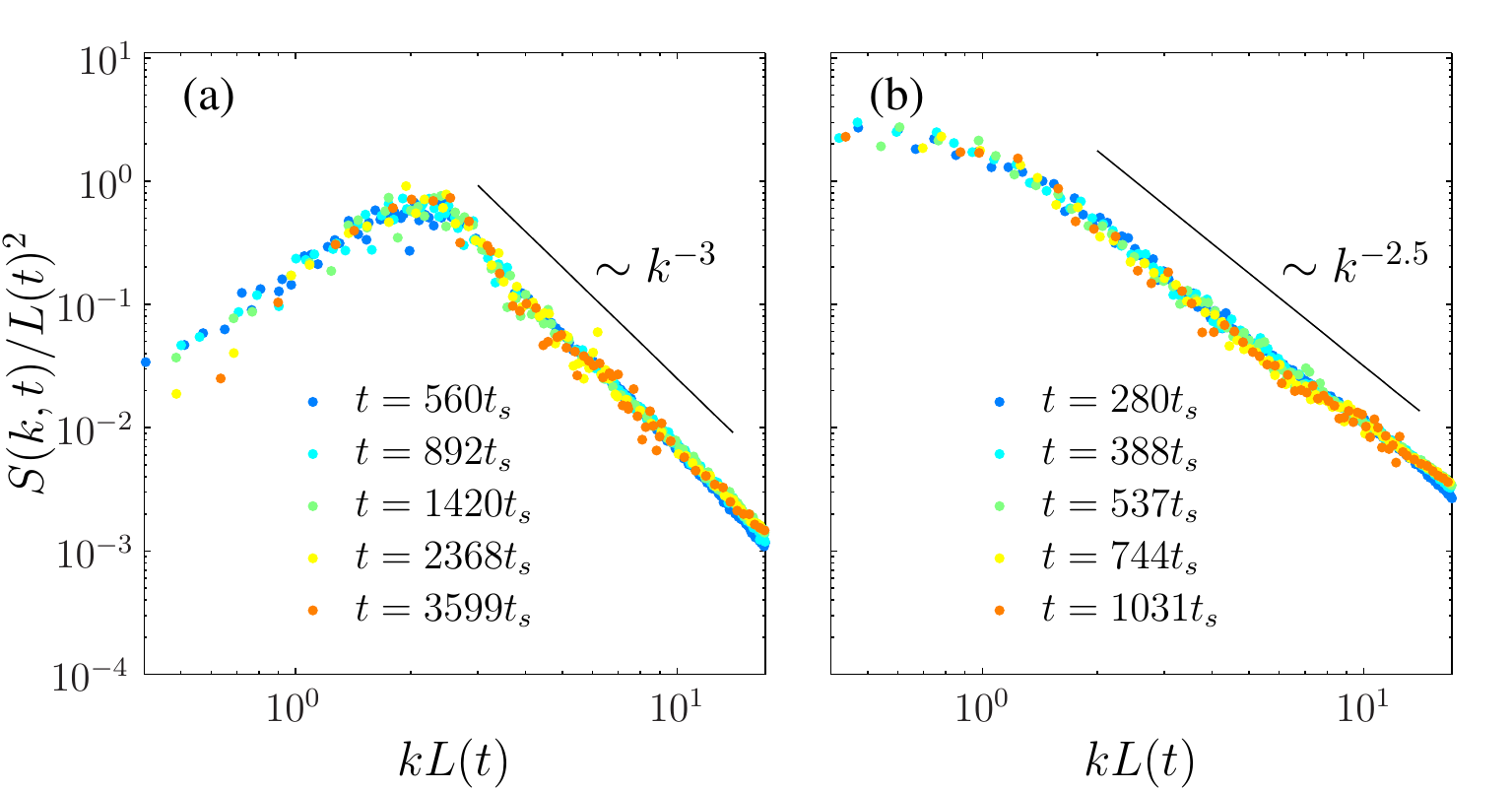}
\caption{\label{SFfig}Plots of structure factor for (a) the easy-axis phase and (b) the easy-plane phase. For $kL>1$ the plots exhibit a longwavelength Porod tail. The $k^{-3}$ tail in the easy-axis phase indicates the presence of smooth domain boundaries, whereas the $k^{-2.5}$ tail in the easy-plane phase is indicative of a fractal domain boundary.}
\end{figure}

In principle it is possible to extract the small $r/L$ behaviour directly from the correlation functions in Figs.~\ref{GLfig}(a) and~(b). Equivalently, we instead choose to examine the order parameter structure factor, which is the Fourier transform of the correlation function,
\begin{align}
S(\bm{k},t)=\int d^2\bm{r}\, G(\bm{r},t)e^{i\bm{k}\cdot\bm{r}}=L^2\hat{f}(kL(t)).
\end{align}
The scaling form follows from setting $G(r,t)= f\left(r/L(t)\right)$, with $\hat{f}$ being the Fourier transform of $f$. Behaviour of the correlation function at length scales $r<L(t)$ then appears at high wavenumbers $kL(t)>1$. Fourier transforming the result~\eqref{bccorr} results in a high wavenumber ``Porod tail'' in the structure factor~\cite{bale1984,Schaefer1986,Sorensen2001a},
\begin{align}\label{SFfrac}
S(k)\sim k^{-2D+D_b}.
\end{align}
Results for the structure factor for the easy-axis and easy-plane quenches are shown in Fig.~\ref{SFfig}(a) and (b), respectively. For the easy-axis case we observe a ``knee'' in the structure factor at $kL\sim 1.3$ followed by a Porod tail $S(k)\sim k^{-3}$ for $L> k^{-1}\gg \xi_s$ that indicates the presence of smooth domain walls i.e. $D_b=1$. We also observe a Porod tail for the easy-plane case, but with a non-integer exponent, $S\sim k^{-2.5}$. This suggests that the easy-plane domain boundaries are fractal, with a dimension of $D_b\approx 1.5$.

\begin{figure}
\includegraphics[width=0.5\textwidth]{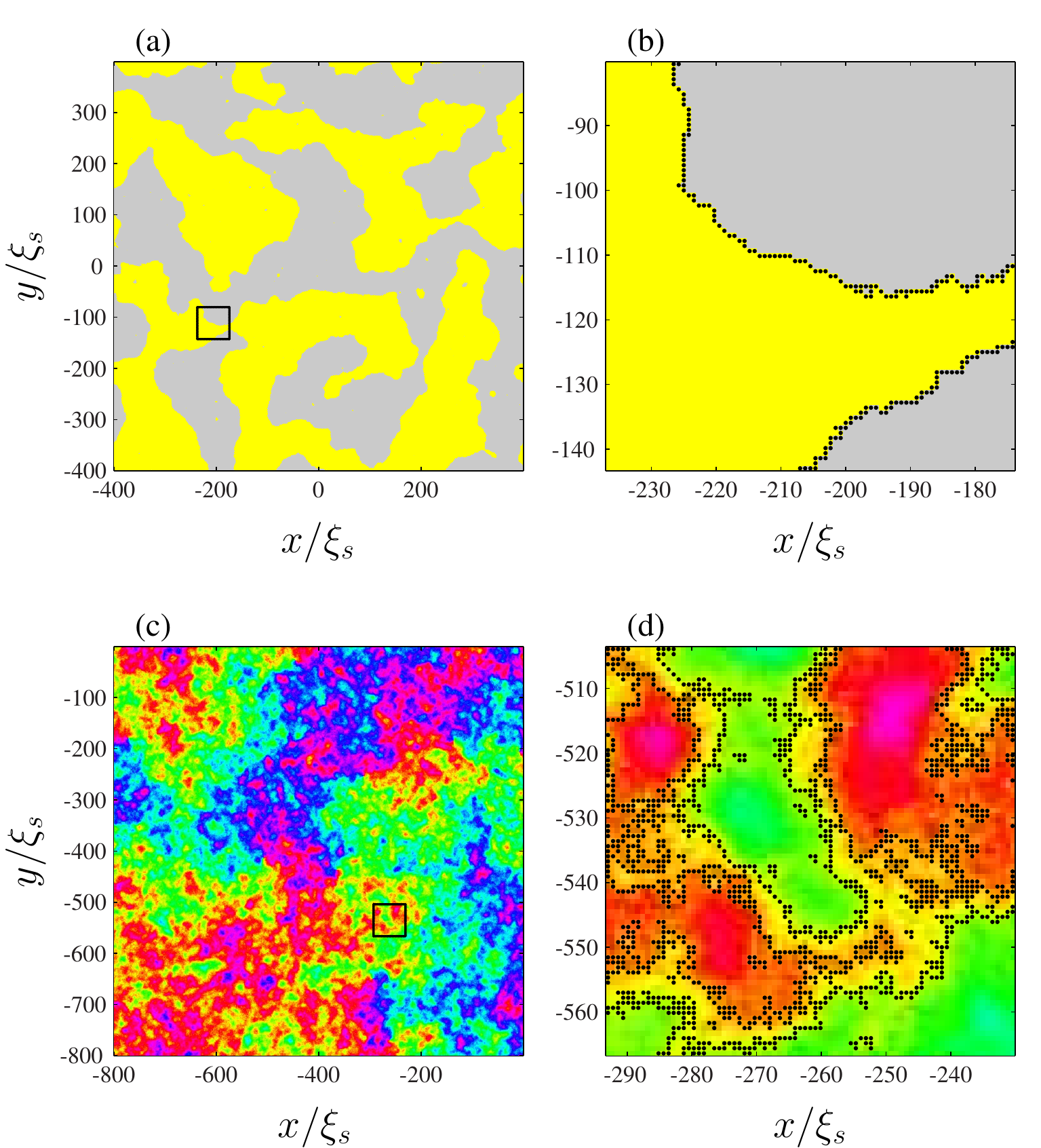}
\caption{\label{fractalFig}Order parameter domains (a),(b) with the boxed region enlarged (c),(d) to show details of domain boundaries. The boundaries are marked by black dots. The axes scale in images (c) and (d) are the same, indicating that the easy-plane domain boundary has a much more convoluted structure than the smooth easy-axis domain boundary. For the easy-axis (easy-plane) phase results displayed are for $q=-0.3q_0$ ($q=0.3q_0$) and $t=3770t_s$ ($t=1131 t_s$).}
\end{figure}

\begin{figure}
\includegraphics[width=0.5\textwidth]{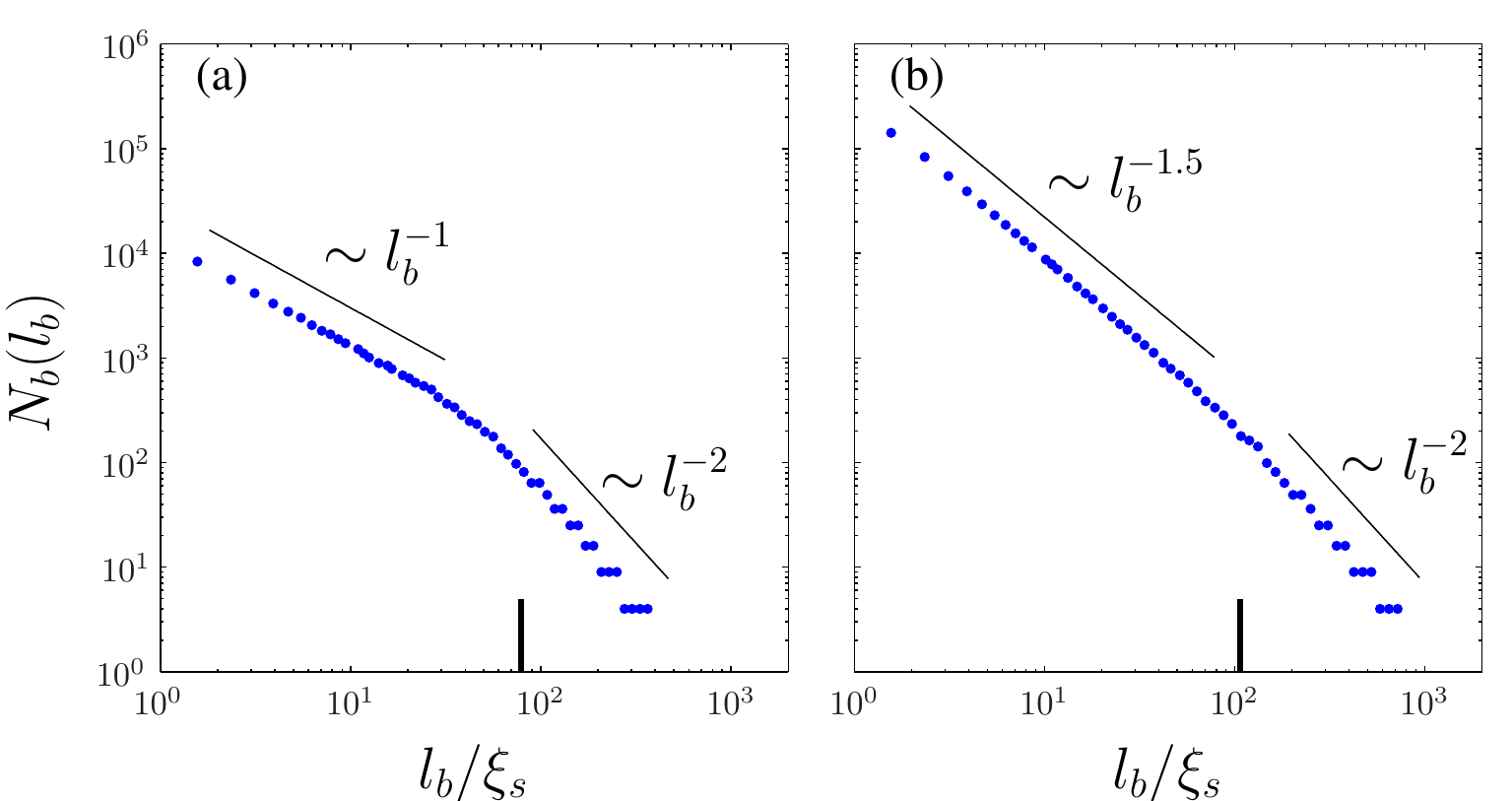}
\caption{\label{boxcountFig}Results for $N_b$ versus $l_b$ obtained from the box-counting algorithm. In the small $l_b$ limit, the plots reveal a box-counting dimension for the domain boundaries of (a) $d_b=1.0$ for the easy-axis phase and (b) $d_b=1.5$ for the easy-plane phase. The vertical lines on the horizontal axes mark the size of domains extracted from the spatial correlation function of the order parameter. For $l_b\gtrsim L(t)$, the slope changes to 2, since boxes larger than the domain size will likely contain a domain boundary. For the easy-axis (easy-plane) phase results displayed are for $q=-0.3q_0$ ($q=0.3q_0$) and $t=3770t_s$ ($t=1131 t_s$).}
\end{figure}

To provide further evidence for this result, we determine a box-counting dimension for the domain boundaries directly. The box-counting dimension is defined through
\begin{align}
d_b=-\lim_{l_b\rightarrow 0}\frac{\log N_b}{\log l_b}.
\end{align}
Here we cover the system with boxes of side length $l_b$ and count the number of boxes $N_b$ that contain a domain boundary. In the limit of small $l_b$ the slope of $\log N_b$ versus $\log l_b$ gives the box-counting dimension. This naturally connects with the probabilistic interpretation of the dimension we used to discuss result~\eqref{bccorr}. In the easy-axis phase the domain boundaries can be identified by looking for non-zero gradients in the sign of $F_z$, see Fig.~\ref{fractalFig}(a),(b). For the easy-plane phase, a single domain is not as well defined because the order parameter changes continuously. However, the domain patterns in Fig.~\ref{domain}(b) do show clear regions of largely one colour. To extract boundaries between these regions, we choose a $\pi/5$ range of (in-plane) spin directions and define discrete domains of spins that lie in this range. By assigning a 1 to spins within the domain and a -1 to spins outside the domain we can identify domain boundaries in an analogous way to the easy-axis phase, see Fig.~\ref{fractalFig}(c),(d). Once domain boundaries have been identified, we can perform a box counting algorithm to determine the box-counting dimension of the boundaries. We do this over an order of magnitude of box sizes, which yields a box counting dimension of $d_b=1.0$ for the easy-axis domain boundaries, see Fig.~\ref{boxcountFig}(a). In comparison, we obtain a box counting dimension of $d_b\approx 1.5-1.6$ for the easy-plane domain boundaries, Fig~\ref{boxcountFig}(b). For the easy-plane phase, we can repeat the box-counting algorithm for different domains of spin range, which give results consistent with Fig.~\ref{boxcountFig}(b). Both the easy-axis and easy-plane box counting dimensions agree with the slopes of the Porod tails in Figure~\ref{SFfig}.  We note that the Porod tail in the easy-plane phase is not accounted for by topological defects (polar-core spin vortices), which would result in a $k^{-4}$ tail \cite{Bray1991a,Bray1994,Rojas1999a}. 

\subsection{Analytic models of coarsening}
Simple analytic models can be used to obtain the dynamic critical exponents found numerically in the previous section. The models describe a dynamic process that is expected to be important in the coarsening dynamics. Imposing dynamic scale invariance on the equation describing this process allows one to extract a dynamic critical exponent.
\subsubsection{Easy-axis}\label{SecEAmodel}
The dynamic critical exponent obtained in the easy-axis phase suggests that inertial hydrodynamics is important in the coarsening~\cite{Furukawa1985,bray2003}. We derive this inertial hydrodynamic process by considering the hydrodynamic formulation of a spin-1 condensate~\cite{Yukawa2012}. Ferromagnetic condensates support both mass and spin superfluid currents. We assume a condensate with constant number density $n$ and zero population in the $m=0$ spin level. Note that these conditions still allow for spatial variation in $|F_z|$, as long as the sum of populations in the $m=\pm 1$ levels is constant. We assume identical phase profiles in the $m=\pm 1$ levels. The $F_z$ superfluid current is then
\begin{align}\label{vFz}
\bm{v}_{F_z}\equiv &\frac{\hbar}{2Mni}\left(\bm{\psi}^\dagger f_z\left(\nabla\bm{\psi}\right)-\left(\nabla\bm{\psi}\right)^\dagger f_z\bm{\psi}\right) 
=\frac{F_z}{n}\bm{v},
\end{align}
where $\bm{v}$ is the mass superfluid velocity,
\begin{align}
\bm{v}\equiv \frac{\hbar}{2Mni}\left(\bm{\psi}^\dagger\left(\nabla\bm{\psi}\right)-\left(\nabla\bm{\psi}\right)^\dagger\bm{\psi}\right).
\end{align}
The current $\bm{v}_{F_z}$ can be understood through the continuity equation (valid when $F_z$ is conserved),
\begin{align}
\frac{\partial F_z}{\partial t}+\nabla\cdot n\bm{v}_{F_z}=0.
\end{align}
Eq.~\eqref{vFz} shows that the order parameter is transported by the mass current. The equation of motion for $\bm{v}$ is
\begin{align}\label{vspindce}
\frac{\partial\bm{v}}{\partial t}+\left(\bm{v}\cdot\nabla\right)\bm{v}=-\frac{g_s}{2Mn}\nabla F_z^2.
\end{align}
We have omitted third order derivative terms in Eq.~(\ref{vspindce}). The scaling properties of our system arise from gradual, large length scale processes for which third order derivatives will be small. Equation~\eqref{vspindce} takes the form of the Euler equation from fluid dynamics, with the term $g_s F_z^2/2Mn$ in place of the pressure. This term is in fact a pressure, as can be seen by considering the energy of regions $A$ across which $F_z$ changes little
\begin{align}
E(A)=\frac{g_n}{2} n^2A+qnA+\frac{g_s}{2}F_z^2A.
\end{align}
The pressure in such a region is
\begin{align}
P(A)=&-\left(\frac{\partial E}{\partial A}\right)_{N,M_z}
\!=\,\frac{g_n}{2} n^2+\frac{g_s}{2} F_z^2
\end{align}
where the partial derivative is evaluated for fixed atom number and total $z$ magnetization $M_z=\int d^2\mathbf{x}\,F_z$. With constant number density the gradual spatial variation of $P(A)$ arises only from the term $g_s F_z^2/2$. Equation~\eqref{vspindce} can therefore be written as
\begin{align}\label{vspindce2}
\frac{\partial \bm{v}}{\partial t}+\left(\bm{v}\cdot\nabla\right)\bm{v}=-\frac{1}{Mn}\nabla P.
\end{align}
In mechanical equilibrium, the pressure difference, $\Delta P$, across a curved surface is related to the surface tension, $\sigma$, of the surface through the \emph{Young-Laplace Equation}~\cite{Scheaybroeck2008} 
\begin{align}\label{YLE}
\Delta P\sim\frac{\sigma}{R},
\end{align}
where $R$ is the curvature of the surface. This relationship arises because the excess surface energy in a curved surface gives rise to a force on this surface, and in equilibrium this must be balanced by a pressure difference. In the case of a condensate with two components separated by a domain wall, the surface energy arises from the kinetic energy across the wall~\cite{Ao1998}. We expect the balance of surface tension and pressure to be fast in comparison to the slow hydrodynamics driving domain growth, so that we can use Eq.~\eqref{YLE} in Eq.~\eqref{vspindce2}.

We now want to impose dynamic scale invariance on Eq.~\eqref{vspindce2}. We carry out a scaling transformation of this equation, rescaling times by $T$, i.e. $t\rightarrow t/T$, and lengths by $L(T)$. We note that $\Delta P$ scales as $L$ [from Eq.~\eqref{YLE}] so that $\nabla P$ scales as $L^2$. This gives
\begin{align}
\frac{T^2}{L}\frac{\partial\bm{v}}{\partial t}+\frac{T^2}{L}\left(\bm{v}\cdot\nabla\right)\bm{v}=-L^2\frac{1}{Mn}\nabla P,
\end{align}
Imposing dynamic scale invariance amounts to Eq.~\eqref{vspindce2} being preserved under this transformation. This can only occur if $L(T)=T^{2/3}$. To write this in the usual scaling form $L(t)\sim t^{1/z}$ we choose $T\propto t$, giving $z=3/2$~\cite{Furukawa1985,Bray1994,Hofmann2014}.

\subsubsection{Easy-plane}\label{SecEPmodel}

For systems in the model E dynamic universality class, the dynamics of spin-waves dynamically coupled to a second conserved field are argued to be the important process during coarsening~\cite{Hohenberg1977,Nam2011,Tauber2014}. Here \emph{dynamically coupled} means that there is no direct coupling between these two fields in the Hamiltonian, but there is a non-vanishing Poisson bracket relation between them that leads to coupling in the dynamic equations.

To show how spin waves arise in our system we begin with a variational ansatz~\cite{Barnett2011}
\begin{align}\label{spinwAnsatz}
\bm{\psi}(\bm{x},t)=\sqrt{n_0}\left(\begin{array}{c}\sin\beta e^{-i\theta}\cos\left(\pi/4+\chi\right) \\\cos\beta\\\sin\beta e^{i\theta}\sin\left(\pi/4+\chi\right)\end{array}\right),
\end{align}
with $\cos(2\beta)=q/q_0$ and variational parameters $\theta(\bm{x},t)$ and $\chi(\bm{x},t)$. In the ground state we would have $\chi=0$ and uniform $\theta$. Spatial variation in $\theta$ corresponds to fluctuations in the direction of transverse spin and will give rise to spin waves. Non-zero $\chi$ describes fluctuations in the (conserved) $F_z$ magnetization\footnote{In the ansatz~\eqref{spinwAnsatz}, we have ignored gapped modes and modes with a steep spectrum ($E_k\sim c_n\hbar k$, with $c_n=\sqrt{g_nn_0/M}$)~\cite{symes2014}. Long wavelength gapped modes will have higher energy and therefore faster dynamics than the gapless modes in~\eqref{spinwAnsatz}. The gapped modes will therefore be less important in the slow coarsening dynamics. Modes with a steep spectrum will also have higher energy at any given $k$, and lower occupation.}. 
\begin{widetext}
We assume fluctuations of $\chi$ are small and so consider fluctuations up to quadratic order in $\chi$ only. The Hamiltonian then takes the form
\begin{align}\label{Hfluc}
H = \int d^2\bm{x}\,&\left[\frac{\hbar^2 n_0\sin^2\beta}{2M}\left(\left|\nabla\theta\right|^2+\left|\nabla\chi\right|^2\right) 
 -2 g_s n_0^2\sin^2\beta\cos(2\beta)\chi^2\right]. 
\end{align} 
The Lagrangian is obtained through a Legendre transformation of the Hamiltonian with respect to the conjugate variables $\bm{\psi}$ and $i\bm{\psi}^\dagger$. This gives $L=i(\bm{\psi}^\dagger\dot{\bm{\psi}})-H$~\cite{garcia1996,Barnett2011}. To second order in $\chi$ and $\dot{\theta}$ we obtain
\begin{align}\label{Lag}
L=2n_0\sin^2\beta\int d^2\bm{x}\left[\chi\dot{\theta}-\frac{\hbar^2}{4M}\left(\left|\nabla\chi\right|^2+\left|\nabla\theta\right|^2\right)+g_sn_0\cos(2\beta)\chi^2\right].
\end{align}
\end{widetext}
Formulating the problem in this form gives the conjugate variable relations
\begin{align}\label{Lquad}
\frac{\delta L}{\delta\dot{\theta}(\bm{x})}=2 n_0\sin^2\beta \chi(\bm{x}),\qquad
\frac{\delta L}{\delta \dot{\chi}(\bm{x})}=0.
\end{align}
The first of these relations reflects the dynamic coupling between fluctuations of the direction of the order parameter and fluctuations of the conserved field $F_z$. This relation also reflects that conservation of $F_z$ magnetization is connected with rotational symmetry about the $F_z$ axis. Evaluating Lagrange's equations~\eqref{Lquad}, which decouple by taking a second time derivative, gives 
\begin{align}\label{thetaphi}
\ddot{\theta}=&q_0\cos(2\beta)\frac{\hbar^2}{2M}\nabla^2\theta-\frac{\hbar^4}{4M^2}\nabla^4\theta,\\ 
\ddot{\chi}=&q_0\cos(2\beta)\frac{\hbar^2}{2M}\nabla^2\chi-\frac{\hbar^4}{4M^2}\nabla^4\chi.
\end{align}
The spin-wave fluctuations ($\theta$) and $F_z$ fluctuations ($\chi$) can be expanded in Fourier modes, giving a spectrum
\begin{align}
E_k=\sqrt{\frac{\hbar^2 k^2}{2M}\left(q_0\cos(2\beta)+\frac{\hbar^2k^2}{2M}\right)}.
\end{align}
For the long wavelength excitations ($k^2\ll 2Mq_0\cos(2\beta)/\hbar^2$),   Eq.~\eqref{thetaphi} gives the equation of motion for spin waves
\begin{align}\label{eqmep}
\ddot{\theta}=\frac{q_0\cos(2\beta)\hbar^2}{2M}\nabla^2\theta.
\end{align}
Scaling lengths by $L$ and times by $T$ in Eq.~\eqref{eqmep}, and setting $T\propto t$, gives scaling $L\sim t^{1/z}$, with $z=1$.

We could alternatively approach the coarsening from the perspective of polar-core spin vortices, Eq.~\eqref{pcvstate}. These are not accounted for in the Lagrangian~\eqref{Lag}, which considers quadratic fluctuations only. A simple model for the dynamics of polar-core spin vortices was given in~\cite{Turner2009},
\begin{align}\label{svmodel}
m\ddot{\bm{r}}_i=\sum_j\frac{\hbar^2 n_0\sin^2\beta}{M}\frac{\kappa_i \kappa_j}{\left|\bm{r}_i-\bm{r}_j\right|^2}\left(\bm{r}_i-\bm{r}_j\right),
\end{align}
where $r_i$ denotes the position of spin vortex $i$ with charge $\kappa_i$, $m$ is the so-called mass of a spin vortex and $\sin\beta$ is defined in Eq.~\eqref{spinwAnsatz}. This model arises by noting that a spin vortex is composed of two scalar vortices in the $m=\pm 1$ levels. Scalar vortices in the $m=1$ level interact according to the usual scalar vortex dynamics~\cite{fetter1965,lucas2014,billam2015}, and similarily for vortices in the $m=-1$ level. However, there is also an attractive interaction between the $m=\pm 1$ vortices within a single spin vortex, which arises from the spin-spin interaction in~\eqref{spinH}. This contributes a crucial core energy that leads to the \emph{second} order equation of motion, Eq.~\eqref{svmodel}. This differs from the \emph{first} order equation of motion that arises in scalar vortex dynamics~\cite{lucas2014}, which would give rise to a dynamic critical exponent of $z=2$. For more details on the dynamics of polar-core spin vortices, see~\cite{Turner2009}.

If we assume that Eq.~\eqref{svmodel} is invariant under a rescaling of lengths by $L$ and times by $T\propto t$, we again obtain the exponent $z=1$. However, the interaction between spin waves and polar-core spin vortices will likely change both the spin wave dynamics~\eqref{eqmep} and the vortex dynamics~\eqref{svmodel}. In the $XY$ model, it is necessary to couple the vortices to damping degrees of freedom to justify the logarithmic correction to scaling~\cite{Yurke1993a}. It may be possible to extend Eq.~\eqref{svmodel} to allow coupling to spin waves~\cite{soffer2016}, which may reveal a logarithmic correction to scaling.

\subsection{Thermalisation of excitations in the easy-axis phase}
As the domains coarsen, energy is liberated into excitations on top of the ordered phase. Over time we expect that these excitations will thermalise. The long wavelength coarsening of the order parameter should be slow compared to other thermalisation processes. Therefore by the time the order parameter domains are large, other modes in the system will have thermalised. We can test this thermalisation by examining the population of Bogoliubov modes on top of the ground state. The Bogoliubov modes may depend on the orientation of the ground state order parameter and so may change across the system due to the presence of domains. 
Also when the Bogoliubov modes and the condensate occupy the same $m$ sublevels it is difficult to distinguish between them.  Fortunately for the easy-axis case the condensate only occupies the $m=\pm1$ sublevels, while there is a spin-wave branch that occupies the $m=0$ sublevel and is insensitive to the orientation of the order parameter, i.e.~not affected by the presence of domains~\cite{Uchino2010}. We can therefore determine the population distribution in the $m=0$ mode and compare this to the value expected in equilibrium.

Since we perform classical field simulations of the quench, we expect to see excitations on top of the ground state populated according to the equipartition theorem when the system is in equilibrium. As will be shown below, we find that the temperature of the $m=0$ modes is large compared to their energy, so that the modes are highly occupied and equipartition is valid. For the spin-1 system there are three Bogoliubov branches on top of the ground state~\cite{Uchino2010,symes2014}. If energy is distributed amongst all the modes of these branches according to equipartition, we would obtain a total energy of $3N^2k_BT$, where $N^2$ is the total number of grid points (i.e.~spatial modes) in the numerical simulation, $T$ is the temperature of the modes and $k_B$ is Boltzmann's constant. Equating this to the total energy liberated from the quench [Eq.~(\ref{Eex})], we obtain the temperature
\begin{align}\label{TEApred}
k_BT=\frac{1}{3}n_0\left(\frac{L}{N}\right)^2\left(\frac{q_0}{4}-q\right),
\end{align}

The $m=0$ Bogoliubov modes (spin-waves) have energy~\cite{Uchino2010}
\begin{align}\label{EkEA}
E_k=\epsilon_k-q,
\end{align}
with $\epsilon_k\equiv\hbar^2 k^2/2M$. According to the equipartition theorem, we expect the population of an $m=0$ mode with energy $E_k$ to be
\begin{align}\label{nkep}
n_k=\frac{k_BT}{E_k}.
\end{align}

\begin{figure}
\includegraphics[width=0.5\textwidth]{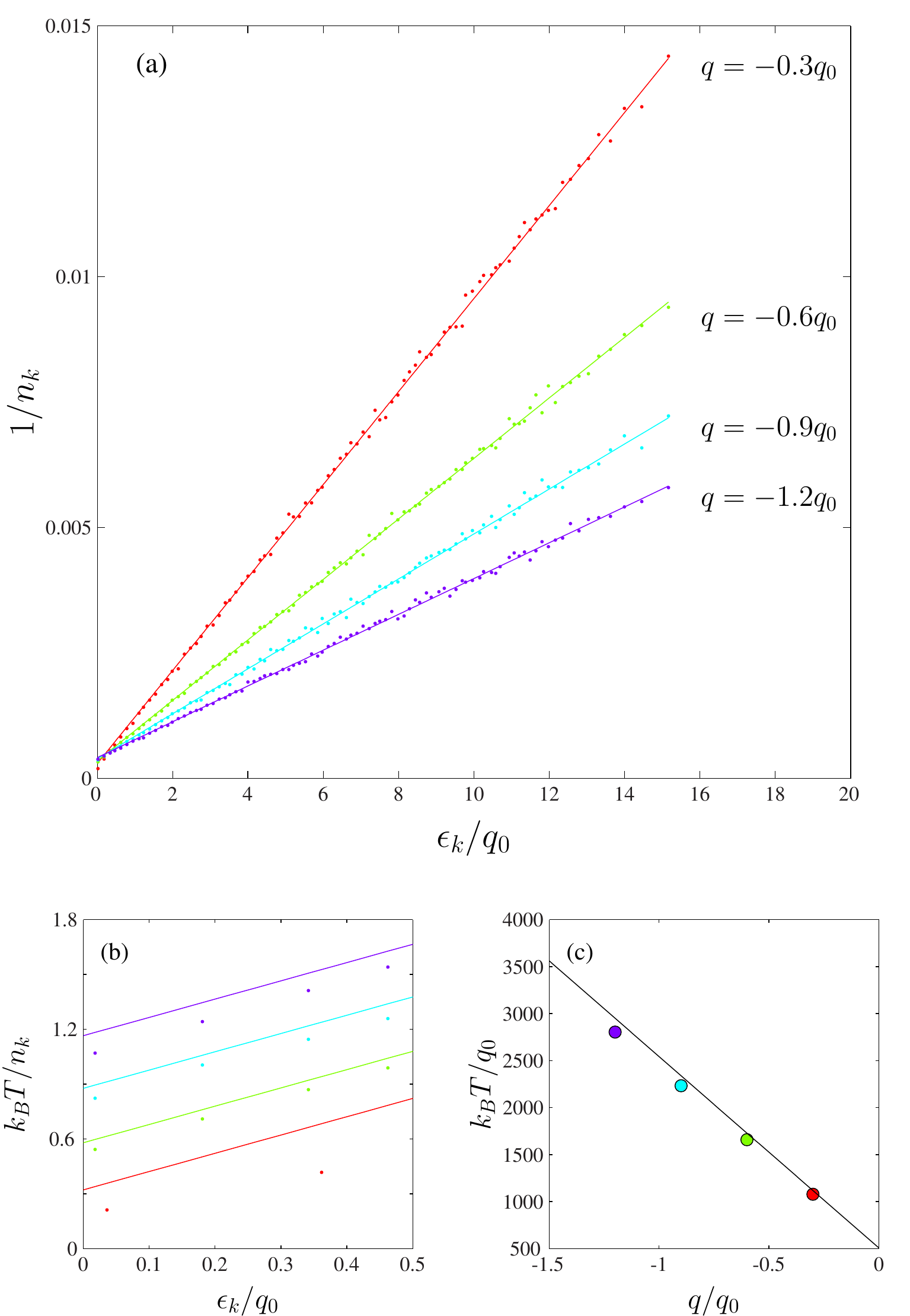}
\caption{\label{EAeqpop}(a) Inverse population of the $m=0$ level across momentum modes for quenches to four different easy-axis $q$ values at time $t=3770t_s$. Dots are numerical data for a single simulation, averaged over azimuthal angle. Lines are best fits of the form $1/n_k=a+bk^2$ where $a$ and $b$ are fitting parameters. (b) Enlarged small $k$ behaviour of (a) (note change of vertical axis). The data agrees well with Eq.~\eqref{nkep}, which predicts that the vertical axis intercept should be $-q/q_0$. (c) Temperature of the $m=0$ excitations versus $q$. Circles are fits from (a), solid line is the the theoretical prediction assuming equipartition, Eq.~\eqref{TEApred}.}
\end{figure}

\begin{figure}
\includegraphics[width=0.5\textwidth]{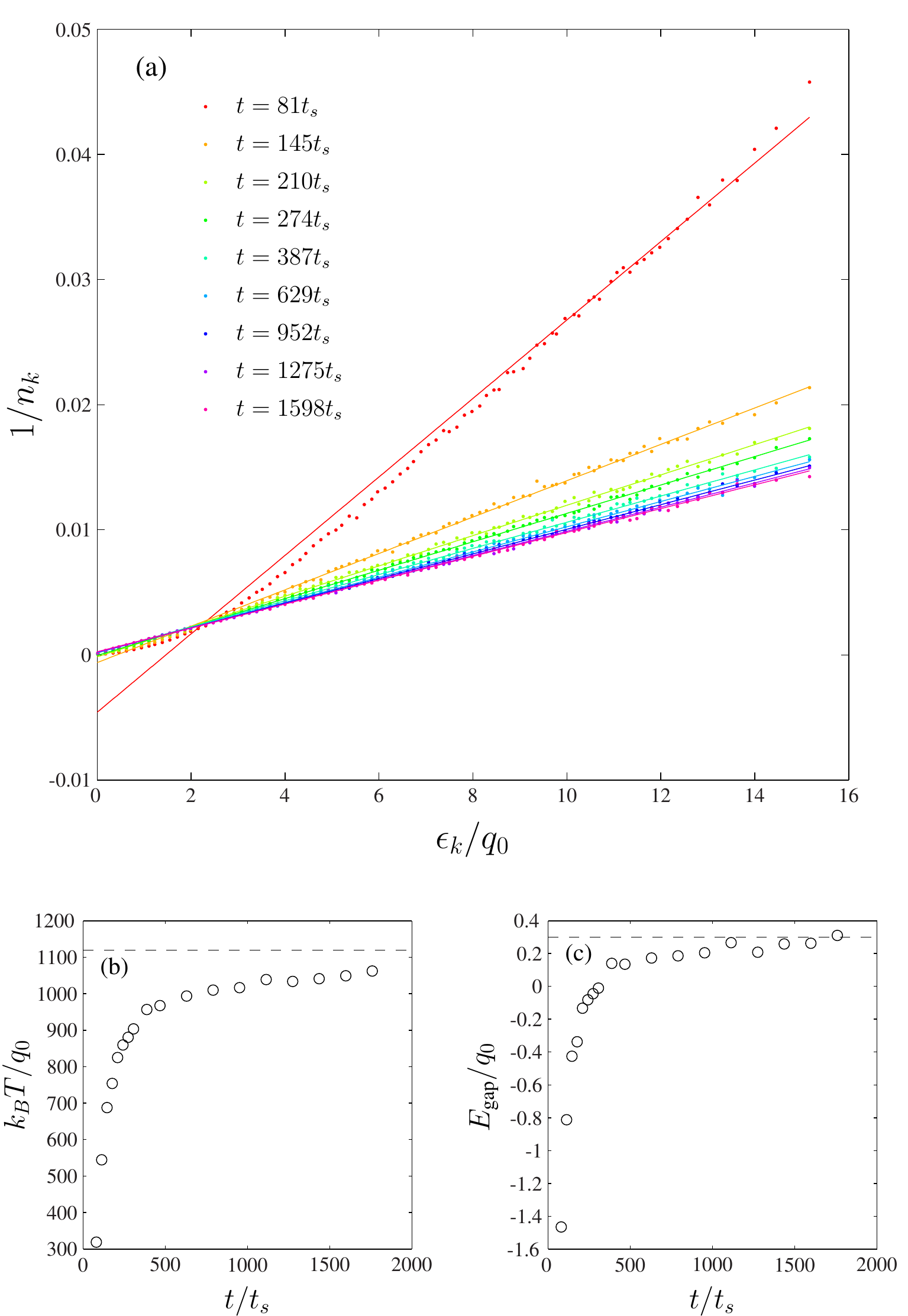}
\caption{\label{EAeqpopt}(a) Inverse population of the $m=0$ level across momentum modes at different times following the quench. Dots are numerical data for a single simulation, averaged over azimuthal angle. Lines are best fits of the form $1/n_k=a+bk^2$ with $a$ and $b$ as fitting parameters. The linear nature of the data shows that energy is equipartioned amongst the (gapped) free particle modes. The slope decreases with time, reflecting an increase in temperature of the $m=0$ excitations. For times $t\lesssim 80 t_s$ (not shown) energy deviates from being equipartioned, coinciding with times where $\left<|F_z|^2\right><\left<|F_\perp|^2\right>$. Data is for $q=-0.3q_0$. (b) Circles show growth of the temperature of the $m=0$ excitations, extracted from the fit to the numerical data in (a). Dashed line shows temperature computed from~\eqref{TEApred}. (b) Circles show the energy gap in the spectrum extracted from the fit to the numerical data in (a) versus time. At late times the gap approaches the value $-q$, marked by the dashed line, in agreement with Eq.~\eqref{EkEA}.}
\end{figure}

\begin{figure}
\includegraphics[width=0.5\textwidth]{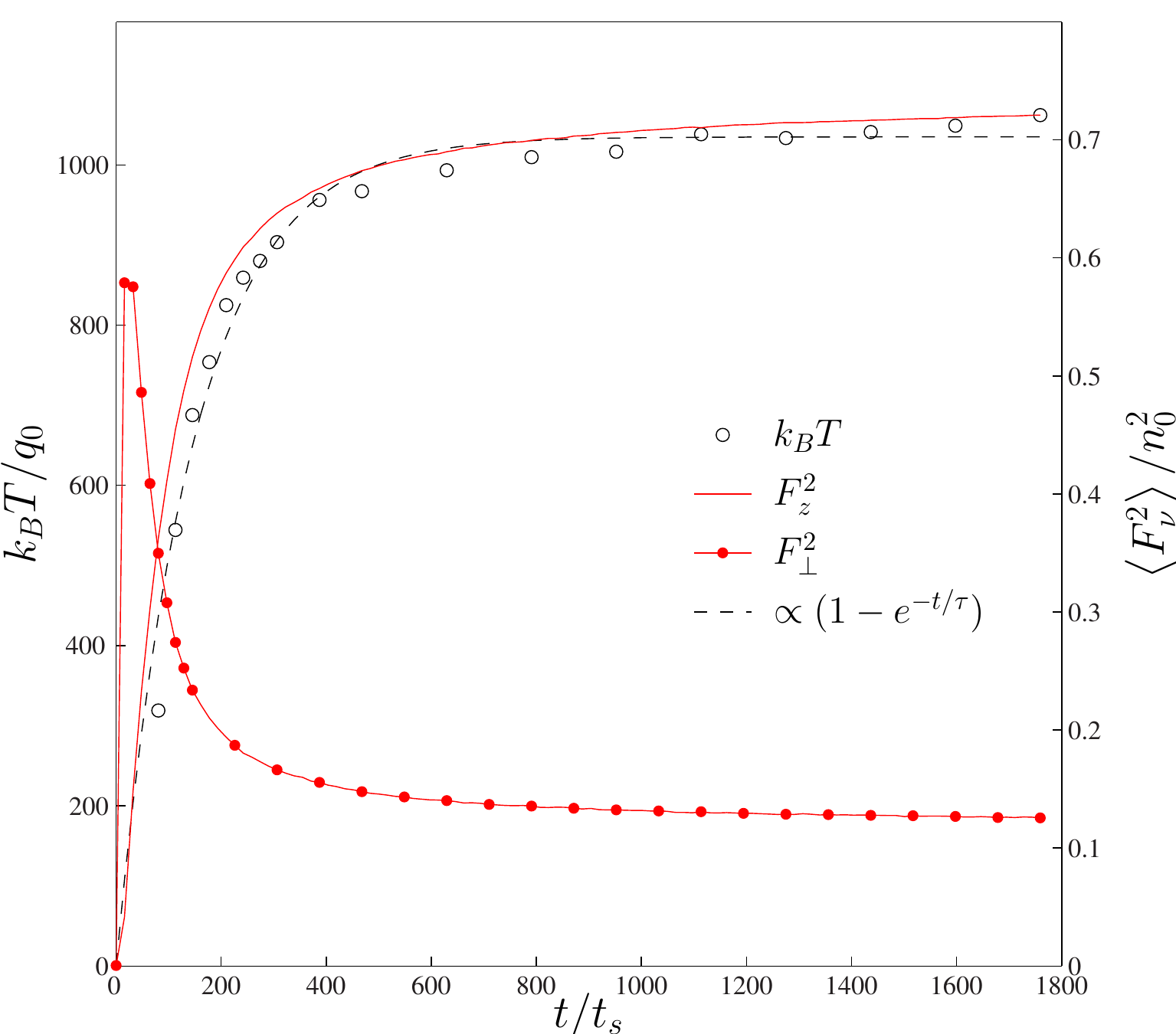}
\caption{\label{growthGT} Growth of temperature (black circles, left axis) and the mean squared $F_z$ magnetization $\left<F_z^2\right>/n_0^2$ (plain red line, right axis) versus time. The black dashed line shows a fit to the temperature growth of the form $\propto (1-e^{-t/\tau})$, where $\tau\approx 150 t_s$ gives the growth rate. The red line with dots shows the decay of the mean squared transverse magnetization $\left<F_\perp^2\right>/n_0^2$ (right axis). Mean squared magnetizations are calculated as described in Fig.~\ref{Allmag}}
\end{figure}

Following Eqs.~\eqref{EkEA} and~\eqref{nkep} we fit late time numerical data for $1/n_k$ to the functional form $a+bk^2$, with $a$ and $b$ as fitting parameters. From $a$ and $b$ we can extract the temperature $k_BT$ and the fitted energy gap $E_\text{gap}$. Figure~\ref{EAeqpop}(a) shows $1/n_k$ data for quenches to $q=-0.3q_0$, $q=-0.6q_0$, $q=-0.9q_0$ and $q=-1.2q_0$. The $a+bk^2$ fits the data well, showing that energy in the $m=0$ level is equipartitioned amongst the momentum modes. The small $k$ behaviour is shown in  Fig.~\ref{EAeqpop}(b) and agrees well with the predicted energy gap of $-q$ [from Eq.~\eqref{EkEA}]. The extracted temperatures are shown in Fig.~\ref{EAeqpop}(c), along with the prediction~\eqref{TEApred}. The agreement between the fitted temperature and predicted temperature is good.

We also determine the rate that energy flows to the $m=0$ Bogoliubov modes by examining how the mode populations change with time, see Fig.~\ref{EAeqpopt}(a). The extracted temperature and energy gap are shown in Figs.~\ref{EAeqpopt}(b) and (c) respectively. Even for quite early times, $t\approx 80 t_s$, energy is equipartitioned amongst the momentum modes. The temperature, however, does not equilibriate until a time $t\sim 500t_s$. The flow of energy to the $m=0$ level is therefore slower than equilibriation amongst the $m=0$ momentum modes. For times $t\lesssim 80 t_s$ (not shown) energy deviates from being equipartitioned, coinciding with times where $\left<|F_z|^2\right><\left<|F_\perp|^2\right>$, see Fig.~\ref{Allmag}(a).

The growth of temperature in Fig.~\ref{EAeqpopt}(c) matches the growth of $F_z$ magnetization in the system, see Fig.~\ref{growthGT}. The time scale of growth also matches the time scale of decay of transverse magnetization. We can determine this time scale by fitting the growth to the functional   form $\propto \left(1-e^{-t/\tau}\right)$. This empirical fit gives a growth rate of $\tau\approx 150 t_s$. We note that the density interaction $g_n n^2$ will likely drive the thermalisation of the $m=0$ momentum modes, whereas the changes in spin density are driven by the smaller spin interaction.

\section{Conclusions and Outlook}
In this paper we have examined the quench dynamics of a quasi-two-dimensional ferromagnetic spin-1 condensate. We have found that for quenches to $q<0$ (into the easy-axis ferromagnetic phase), order grows with a dynamic critical exponent of $z=3/2$ (also see \cite{Kudo2013a,Mukerjee2007a,Hofmann2014}). For quenches to $q$ in the range $0<q<q_0$ (into the easy-plane ferromagnetic phase) we find that order grows with a dynamic critical exponent of $z=1$. With our numerical results we have verified that the late time coarsening dynamics is scale invariant by demonstrating correlation function collapse when scaling by $L(t)$. For the easy-plane quench we demonstrate the important role of polar-core vortices by showing that the number of vortices scales as $L(t)^{-2}$ during the coarsening dynamics. Thus we can interpret easy-plane coarsening as occurring via the mutual annihilation of vortex anti-vortex pairs.

To provide insight into the origin of the growth of order we have discussed simple analytic models that capture the essential dynamics of the coarsening, and correctly predict the dynamic critical exponents. For the easy-plane phase we can develop such a model based either on the spin-waves or the vortex dynamics, although are unable to analytically obtain the log corrections to the growth law, which remains an area for future investigation.

The structure of the ordered domains was studied by quantifying the Porod tails in the order parameter structure factor, and by a direct spatial analysis using the box counting algorithm. Both approaches show that the easy-axis domain walls are regular with a dimension of $D_b=1$, whereas the easy-plane domain walls are fractal with a dimension of $D_b\approx1.5$. Possible physical implications of this fractal structure includes diffusion limited aggregation \cite{Halsey2000}, or Schramm (stochastic)-Loewner evolution and the associated conformal invariance \cite{Cardy2005}. We also note recent work considering the domain size distribution and domain wall percolation in binary condensates \cite{Takeuchi2015a,Takeuchi2016a}, which would be an interesting direction to pursue for the spinor system.

We have also considered how the energy liberated from the quench rethermalizes in the system. The deeper the quench more energy is liberated and the system exhibits larger fluctuations. In order to quantify the thermalization in the easy-axis quench we demonstrate an analysis technique that allows us to measure the temperature and distribution of spin-waves.
Using this we show that local equilibrium is established in the spin waves on moderate time scales, but continues to evolve as the order parameter domains anneal.

Finally we discuss the requirements that must be met to observe coarsening in experiments. First, the the size of the condensate must be much larger than the spin healing length $\xi_s$.  With the development of flat optical traps (see \cite{Chomaz2015a,Navon2015a}) this condition is easily met. Furthermore, these flat traps minimise inhomogeneous effects present in harmonic trapping potentials, and appear ideally suited to studies of phase transition dynamics. The second requirement is that the system has a sufficiently long lifetime that the coarsening dynamics can be monitored over timescales much longer than the spin time $t_s$. In $^{87}$Rb this time is typically $t_s\sim100\,$ms and experiments have been able to study coarsening dynamics for times up to $4\,$s \cite{Guzman2011a}. Our results here would suggest that timescales up to an order of magnitude longer would be necessary to observe universal coarsening behaviour.

\section{Acknowledgments}
 We gratefully acknowledge support from the Marsden Fund of the Royal Society of New Zealand.

\end{document}